\documentclass[10pt,aps,prc,showpacs,superscriptaddress,twocolumn,preprintnumbers,floatfix,showkeys]{revtex4-1}

\usepackage{amssymb}
 \usepackage{amsmath}
\usepackage[figuresright]{rotating}
\usepackage{slashed}

\newcommand{\La}{{\Lambda}}
\newcommand{\Si}{{\Sigma}}

\newcommand{\be}{\begin{eqnarray}}
\newcommand{\ee}{\end{eqnarray}}

\newlength{\feynwidth} 
\newlength{\feynwidthbig} 

\usepackage{xcolor} 

\begin{document}



\title{Exploring the $\Lambda$-deuteron interaction via correlations in heavy-ion collisions}
\author{J. Haidenbauer}
\affiliation{Institute for Advanced Simulation, Institut f\"ur Kernphysik (Theorie), and 
J\"ulich Center of Hadron Physics, Forschungszentrum J{\"u}lich, D-52425 J{\"u}lich, Germany}
\begin{abstract}
$\Lambda$-deuteron two-particle momentum correlation functions, to be 
measured in high-energy heavy-ion collisions, are investigated. In particular, 
the question is addressed whether such correlations can serve as an additional and 
alternative source of information on the elementary $\Lambda N$ interaction.
The study is performed within the Lednicky-Lyuboshits formalism, 
utilizing an effective range expansion for the two relevant $S$-wave $\Lambda d$ 
amplitudes with parameters taken from the literature.   
It is found that in collisions characterized by a large emitting source
the $\Lambda d$ correlation function is predominatly sensitive to the 
quartet state ($^4S_{3/2}$). In contrast, for small source sizes the contribution 
from the doublet partial wave ($^2S_{1/2}$) could be significant. Though the 
latter is constrained by the hypertriton binding energy, its present experimental 
uncertainty impedes an accurate determination of the doublet amplitude and, 
in turn, complicates conclusions on the quartet state.
\end{abstract}

\maketitle

\section{Introduction}

Contrary to the nucleon-nucleon ($NN$) interaction, the forces between
strange baryons ($\La$, $\Si$, $\Xi$) and nucleons are still poorly understood, 
not least because of the limited number and accuracy of available scattering data
\cite{Engelmann:1966,Alexander:1968,SechiZorn:1968,Kadyk:1971}. 
At least with regard to the $\La N$ interaction the main features are roughly 
known due to the aforementioned scattering data but also from measurements 
and studies of hypernuclei \cite{Gal:2016,Botta:2017}. However, there is 
no explicit information on the spin dependence, necessary to resolve the relative 
strength of the interactions in the two possible spin configurations, $S=0,\, 1$.
With respect to that, few-body systems constitute a valuable 
complementary source of information \cite{Gibson:1994}. 
In particular, this concerns the bound systems $^{\La}_3$H (hypertriton) 
and the four-body states $^{\La}_4$H and $^{\La}_4$He, where empirical values for 
the binding energies have been available for a long time already \cite{Juric:1973}. 
Light systems are amenable to a treatment within, e.g., the Faddeev/Yakubovsky 
approach \cite{Miyagawa:1993,Miyagawa:1995,Nogga:2002,Nogga:2013}
or via {\it ab initio} calculations based on the no-core shell model 
\cite{Wirth:2014,Gazda:2016,Wirth:2019}, allowing for a rigorous inclusion 
of the underlying $\La N$ interaction and of the important coupling to the 
$\Si N$ system. 
Clearly, there should be also three-body forces (3BF) \cite{Petschauer:2016}, 
which complicate conclusions on the elementary $\La N$ interaction from 
few-body studies. However, since the systems are very light and only 
loosely bound, effects from 3BFs are expected to be small.  Indeed, 
this notion has been adopted by the J\"ulich-Bonn-Munich group in their studies
of the hyperon-nucleon ($YN$) interaction within chiral effective field theory (EFT) 
by considering not 
only the $\La p$ (and $\Si N$) data but also the hypertriton binding energy to 
fix the interaction strength in the spin singlet ($^1S_0$) and triplet ($^3S_1$) 
channels \cite{Polinder:2006,Hai:2013,Haidenbauer:2019,Petschauer:2020}. 

In the present paper we want to explore the potential of an additional and 
independent source of information, namely the $\La$-deuteron ($\La d$) system. 
Certainly, empirical information on direct $\La d$ scattering is even harder to get 
than on, say, $\La p$ scattering, and as far as we know has never been considered. 
However, there is another possibility to access such information, namely by 
means of two-particle momentum correlation functions
\cite{KP1,LL1,KP2,Bauer:1992,Morita:2015,Ohnishi:2016,Cho:2017,Haidenbauer:2018},
measured in heavy-ion collisions and/or high-energetic $pp$ 
collisions. Such correlations were initially considered as a tool to learn more 
about the emission process and/or the properties of the emitting source. 
But they provide likewise a doorway to information on hadron-hadron forces 
at low energies, specifically on those that are inaccessible by other means. 
Experiments with that aim in mind have been suggested and (in part) already 
successfully performed for multistrange systems like 
$\La \La$ \cite{Morita:2015,Ohnishi:2016,Adamczyk:2015,Acharya:2018}, 
$p\Omega$ \cite{STAR:2019,Morita:2016} or $\Omega\Omega$ \cite{Morita:2020},
and also for charmed baryons \cite{Haidenbauer:2020}.
Extending the measurments to $\La d$ correlations could be feasible too,
judging from the available production yields \cite{Braun-Munzinger:2019}. 
In fact, in the past, experimental studies of correlations for $p d$, 
$d d$ and even for light nuclei have been already performed 
\cite{Chitwood:1985,Pochodzalla:1987,Kotte:1999,Wosinska:2007,Aggarwal:2007}, 
and a measurment of $K^- d$ correlation functions is in progress
\cite{ALICE:Kd}. Actually, even $\La d$ has been on the 
agenda \cite{Wisniewski:2012}. 

Whereas calculations of the hypertriton abound in the literature
there is little to be found about $\La d$ scattering. 
This is not too surprising, since, as said before, the prospects of pertinent 
scattering experiments are practically non-existent. Nonetheless, there is 
a series of Faddeev-type studies and also variational calculations starting with 
the pioneering work of Schick and Collaborators in the 1960s
\cite{Schick:1965,Schick:1967,Schick:1969,Schick:1971}
and followed by others 
\cite{Kolesnikov:1982,Hasegawa:1983,Afnan:1993,Cobis:1997,Filikhin:2000,Garcilazo:2007,Garcilazo:2007D}.
Recently,  $\La d$ scattering at energies close to the threshold has been 
studied within pionless effective field theory ($\slashed{\pi}$EFT)
\cite{Hammer:2002,Hildenbrand:2019,Schafer:2020}.

The present work is intended to serve as illustration for what can be expected from 
measuring $\La d$ correlations. It is an exploratory study and, therefore, it is done 
on a simple technical level.
For the interpretation of actual data on $\La d$ correlation functions it is certainly
advisable to perform solid and full-fledged calculations of the $\La d$ system. 
It goes without saying that such calculations are challenging and technically demanding.
%
Since the hypertriton is weakly bound and the binding energy is known \cite{Juric:1973}
(see, however, Refs. \cite{Adam:2019,Le:2019}), 
effective range theory can be used to pin down the $\La d$ $S$-wave amplitude in the 
spin-doublet state ($^2S_{1/2}$) at low energies in an essentially model-independent 
way \cite{Hammer:2002,Bethe:1949}. 
The situation is much less satisfactory for the spin-quartet ($^4S_{3/2}$) amplitude
as demonstrated by the results reported in Ref. \cite{Schafer:2020}.
Thus, the essential question to be addressed is in how far a measurement of the 
$\La d$ correlation function could help to pin down the latter amplitude. 

The paper is structured in the following way: In Sect. \ref{Sec:Formalism}
the formalism for two-particle momentum correlation functions is briefly 
reviewed. Specifically, the simple and compact expression for the correlation
function due to Lednicky and Lyuboshits \cite{LL1} is provided, which is used 
for the present investigation. 
Results for the $\La d$ correlation functions are presented in 
Sect. \ref{Sec:Results}. A variety of amplitudes for the $^2S_{1/2}$ and $^4S_{3/2}$
partial waves is considered, all taken from the literature, and their influence
on the resulting correlation functions is discussed. In addition the role of 
the size of the emitting source (parameterized in terms of a Gaussian source
function) on the results is explored. 
The paper ends with concluding remarks.

\section{Correlation function}
\label{Sec:Formalism}

The formalism for calculating the two-particle correlation function
has been described in detail in various publications
\cite{KP1,LL1,KP2,Bauer:1992,Morita:2015,Ohnishi:2016,Cho:2017}. 
We summerize it here very brief and provide only an overview of the essential formulae. 
The two-particle momentum correlation function is defined by
\begin{align}
C(\bold{p_1},\bold{p_2})
=&\frac{
\int d^4x_1 d^4x_2
S_1(x_1,\bold{p}_1)
S_2(x_2,\bold{p}_2)
\left| \Psi^{(-)}(\bold{r},\bold{k}) \right|^2
}{
\int d^4x_1 d^4x_2
S_1(x_1,\bold{p}_1)
S_2(x_2,\bold{p}_2)
}
\nonumber\\
\simeq&
\int d\bold{r}
S_{12}(\bold{r})
\left| \Psi^{(-)}(\bold{r},\bold{k}) \right|^2 \ .
\label{Eq:Corr}
\end{align}
Here the quantity $S_i(x_i,\bold{p}_i)~(i=1,2)$ is the single particle source function
of particle $i$ with momentum $\bold{p}_i$.
As already indicated by Eq.~\eqref{Eq:Corr}, we evaluate the quantity in question in the
center-of-mass (c.m.) frame where the wave function $\Psi^{(-)}$ is then a function
of the relative coordinate $\bold{r}$ and the c.m. momentum,
$\bold{k}=(m_2\bold{p}_1-m_1\bold{p}_2)/(m_1+m_2)$, 
and $S_{12}(\bold{r})$ is the normalized pair source function that depends likewise
only on the relative coordinate. Furthermore, we consider only interactions in the $S$-wave.

Assuming a static and spherical Gaussian source with radius $R$,
$S(x,\bold{p})\propto \exp(-\bold{x}^2/2R^2)\delta(t-t_0)$,
a partial wave expansion can be performed straightforwardly and the correlation 
function can be written in a compact form \cite{Ohnishi:2016}.
In particular, for systems with two non-identical particles such as $\Lambda p$ or 
$\La d$ the correlation function amounts to
\begin{align}
C(k)\simeq
1+\int_0^\infty 4\pi r^2\,dr\, S_{12}(\bold{r})
\left[
\left|\psi(k,r)\right|^2
-\left|j_0(kr)\right|^2
\right] \ ,
\label{Eq:cni}
\end{align}
where the properly normalized source function is given by 
$S_{12}(\bold{r})=\exp(-r^2/4R^2)/(2\sqrt{\pi}R)^3$ and $j_l(kr)$ is the spherical 
Bessel function for $l=0$. 
$\psi(k,r)$ is the scattering wave function. For two-body systems it 
can be obtained easily by solving the Schr\"odinger equation for a given potential, 
but also from the Lippmann-Schwinger (LS) equation \cite{Haidenbauer:2018}.
In case of $\La d$, in principle, the wave functions can be deduced from the solution 
of the configuration-space Faddeev equations or from variational calculations
\cite{Chen:1989,Kievsky:1994,Gloeckle:1996}. As a less ambitious alternative 
one could construct effective $\La d$ two-body potentials \cite{Congleton:1992}
following corresponding studies for the $N\,d$ case \cite{Tomio:1987,Orlov:2000},
and use them for generating wave functions. 
In general, $S$-wave states of two particle can be formed with different spins.
For example, $\La N$ can be in the partial waves $^1S_0$ and $^3S_1$, respectively,
and $\La d$ in the $^2S_{1/2}$ and $^4S_{3/2}$ states. 
Accordingly, an averaging over the spin has to be performed in Eq.~(\ref{Eq:cni}). 
It is usually assumed that the weight is the same as for free scattering.
For the $\La d$ system the weights are $1/3$ and $2/3$, respectively, i.e.
$|\psi (r,k)|^2 \to \frac{1}{3}|\psi_{1/2}(r,k)|^2 + \frac{2}{3}|\psi_{3/2}(r,k)|^2$.

A much simpler expression for the correlation function can be derived if one assumes 
that the wave function entering Eq. (\ref{Eq:cni}) can be approximated by its 
asymptotic form, $\psi (k,r) \to j_0(kr) + f(k) \exp (ikr)/r$. 
Then one arrives at a formula often called the Lednicky-Lyuboshits (LL) approach 
or model \cite{LL1}:

\begin{eqnarray}
&&\int_0^\infty 4\pi r^2 dr\,S_{12}(r) 
\left[
\left|\psi(k,r)\right|^2
-\left|j_0(kr)\right|^2
\right] 
\approx \nonumber \\
&&
\frac{|f(k)|^2}{2R^2} F(r_0) 
+\frac{2\text{Re}f(k)}{\sqrt{\pi}R}\,F_1(x)
-\frac{\text{Im}f(k)}{R}\,F_2(x) \ . 
\label{Eq:LL}
\end{eqnarray}

Here $f(k)$ is the scattering amplitude which is related to the $S$-matrix by $f(k)=(S-1)/2ik$,
and in practical applications is often replaced by the effective range expansion (ERE), i.e.
$f(k)\approx 1/(-1/a_0 + r_0k^2/2 - i k)$ with $a_0$ and $r_0$ being the scattering length 
and the effective range, respectively. Furthermore, $F_1(x)=\int_0^x dt\, e^{t^2-x^2}/x$ 
and $F_2(x)=(1-e^{-x^2})/x$, with $x=2kR$. 
The factor $F(r_0) = 1-r_0/(2\sqrt{\pi}R)$ is a correction that accounts for the deviation of
the true wave function from the asymptotic form \cite{LL1,Ohnishi:2016}.
The approximation (\ref{Eq:LL}) works reasonably well for source sizes $R$ larger 
than the range of interaction. For smaller values of $R$ there might be noticeable
differences between the results with the LL formula and those based on the full
wave function \cite{LL1,Cho:2017}. 

In the present work we show results for three different $R$ values,  where the choice 
is motivated by values suggested by analyses of measurements of the 
$\La p$ correlation  function in $pp$ collisions at $7$ TeV by the ALICE 
Collaboration ($R=1.2$ fm) \cite{Acharya:2018} 
and that of $p\Omega$ in peripheral and central Au+Au
collisions at 200 GeV by the STAR Collaboration ($R=2.5,\, 5$ fm) \cite{STAR:2019}.
For a general discussion of the dependence of correlation functions on the source size 
in combination with the scattering length see, e.g., Refs. \cite{Cho:2017,Morita:2020}. 

As noted in Ref. \cite{MS:2019}, 
Eq.~(\ref{Eq:cni}) is true only if the deuteron behaves like an ``elementary'' particle, 
in the sense that it is directly emitted from the source. If it is formed afterwards 
then modifications are required as discussed in detail in that reference. 
Specifically, in the latter case there will be a modification of the source size $R$. 
We perform the calculations for fixed $R$ values and, thus, we will not consider 
this effect here which is in the order of 15\% or so \cite{MS:2019}. 
In any case, it is open question which $R$ values to expect for $\La d$ production
in different collisions and at different collision energies! 

\section{$\Lambda d$ scattering and $\Lambda d$ correlation functions}
\label{Sec:Results}

In this exploratory study we calculate the $\La d$ correlation functions in the LL 
formalism (\ref{Eq:LL}), based on $\La d$ ERE parameters taken from the literature.  
Values for the parameters in the doublet $S$-wave
can be found in a variety of works
\cite{Cobis:1997,Filikhin:2000,Garcilazo:2007,Garcilazo:2007D,Hammer:2002,Hildenbrand:2019} .
As mentioned above, there is a bound state in this partial wave, the hypertriton, which 
provided an important incentive for pertinent calculations. Indeed the $^3_{\La}$H 
binding energy is related to the ERE parameters in terms of the Bethe 
formula \cite{Bethe:1949} which reads
\begin{eqnarray}
\frac{1}{a_{1/2}} &=& \gamma - \frac{1}{2}\, r_{1/2}\, \gamma^2 \ .
\label{Bethe}
\end{eqnarray}
Here $\gamma$ is the binding momentum; the
binding energy itself is given by $B_\La = \frac{\gamma^2}{2\mu_{\La d}}$
with $\mu_{\La d}$ being the reduced mass of the $\La d$ system. Since the 
binding energy is experimentally known, $B_\La=0.13\pm 0.05$ MeV \cite{Juric:1973},
and very small, it provides substantial constraints on the ERE parameters
and, in turn, on the contribution of this partial wave to the $\La d$ correlation 
function.

For the present study we consider ERE parameters from three-body calculations 
which predict a $^3_{\La}$H binding energy close to the aforementioned 
value of $0.13$ MeV. This is fulfilled for some of the phenomenlogical  
potential sets considered in Ref. \cite{Cobis:1997} and for the calculation of 
H.-W. Hammer \cite{Hammer:2002} based on $\slashed{\pi}$EFT. 
Actually, in the latter work the binding energy is used as input. The found ERE 
parameters are $a_{1/2}=16.8^{+4.4}_{-2.4}$~fm, $r_{1/2}=2.3\pm 0.3$~fm,
where the errors are due to the uncertainty in the $^3_{\La}$H binding energy. 
With regard to potential models we take the result for the combination 
GC1-E5rb from Cobis et al. \cite{Cobis:1997} (cf. Table 6), i.e.  
$a_{1/2} = 16.3$ fm, $r_{1/2} = 3.2$ fm. 
Taking into account the uncertainty in $B_\La$ leads to 
$a_{1/2} = 16.3^{+4.0}_{-2.1}$ fm.  
Since the calculations in Ref. \cite{Cobis:1997} suggest that $r_{1/2}$ is largely
insensitive to variations of the potentials (the hypertriton binding energy),
we kept here the effective range fixed. We note that both sets of effective range 
parameters are reasonably well in line with Eq. (\ref{Bethe}).

\begin{figure*}
\begin{center}
\includegraphics[height=72mm]{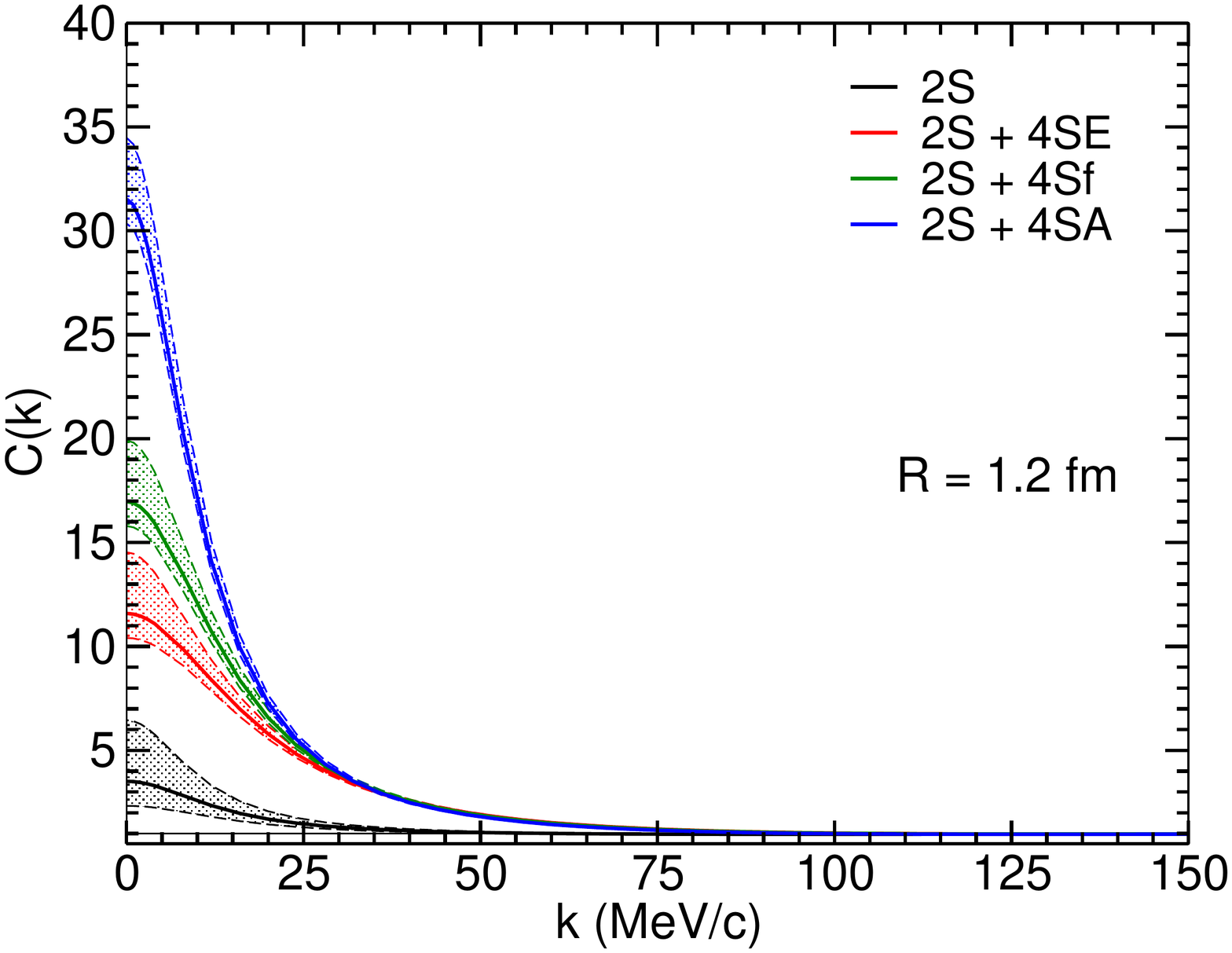}\includegraphics[height=72mm]{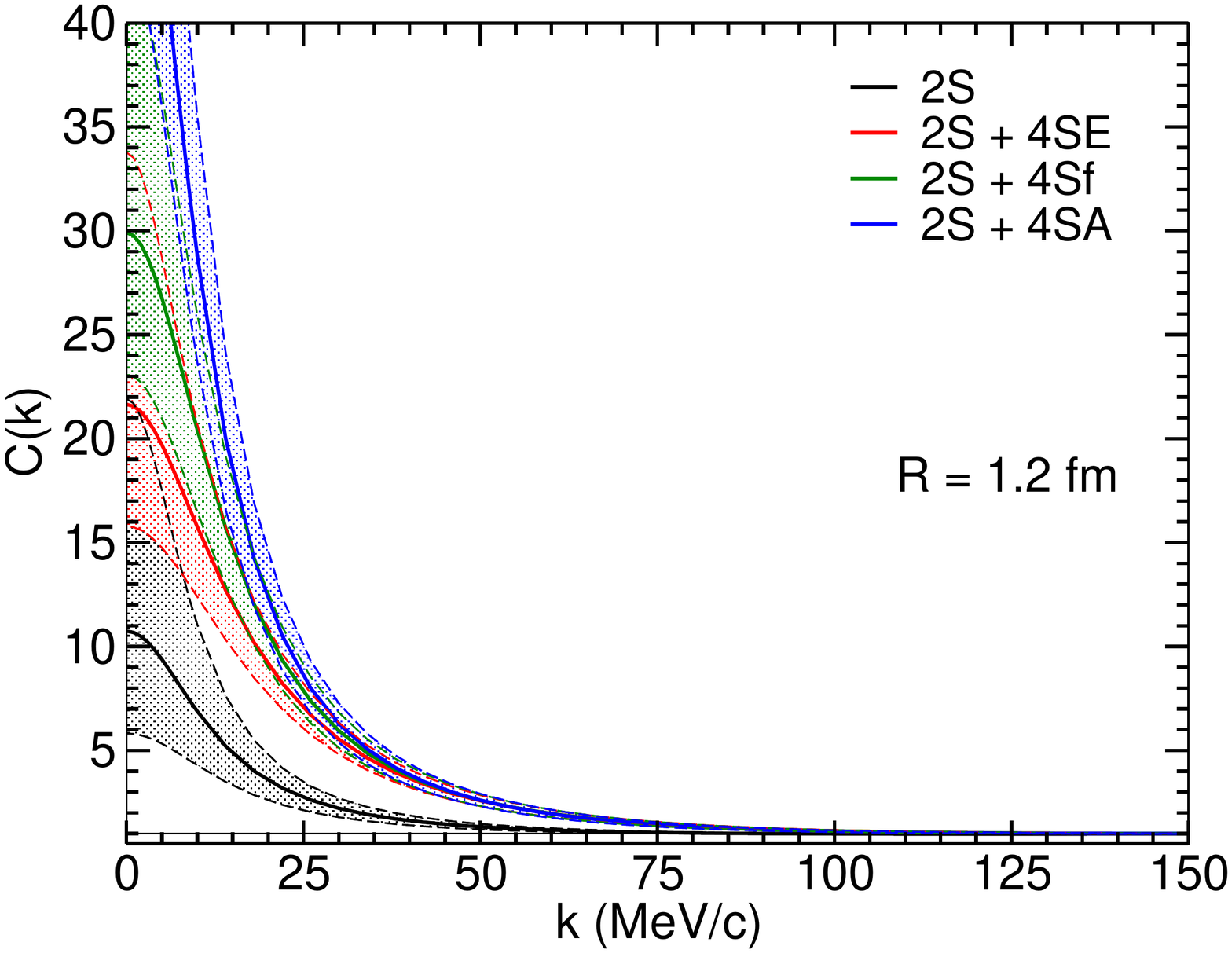}
\caption{
$\Lambda d$ correlation functions for the source size $R=1.2$ fm. Spin-averaged
results are shown where in the $^2S_{1/2}$ state either the ERE parameters
of Cobis \cite{Cobis:1997} (left) or of Hammer \cite{Hammer:2002} (right) are employed. 
For the $^4S_{3/2}$ state results from Sch\"afer \cite{Schafer:2020} are used, 
building on $\La N$ scattering lengths from Alexander (A) \cite{Alexander:1968},
NSC97f (f) \cite{Rijken:1999}, or chiral EFT (E) \cite{Hai:2013} 
(from top to bottom), see text. 
The bands are the error due to the uncertainty in the $^\La_3$H binding energy.
}
\label{fig:Ald12}
\end{center}
\end{figure*}

Results for the ERE parameters in the quartet state are
much harder to find in the literature. We use here the ones from the recent 
calculations of Sch\"afer et al. \cite{Schafer:2020} performed in $\slashed{\pi}$EFT. 
This work reports values (cf. Table II therein) ranging from 
$a_{3/2} = -17.3$ fm, $r_{3/2} = 3.6$ fm, based on a 
$\La N$ interaction fixed by the scattering lengths of 
Alexander et al. \cite{Alexander:1968} 
over $a_{3/2} = -10.8$ fm, $r_{3/2} = 3.8$ fm ($\La N$ properties
adjusted to the Nijmegen $YN$ potential NSC97f \cite{Rijken:1999})
to $a_{3/2} = -7.6$ fm, $r_{3/2} = 3.6$ fm,  with $\La N$ 
fixed by the $YN$ results from a potential derived within SU(3)
chiral EFT up to next-to-leading order 
(NLO13) \cite{Hai:2013}. 
An even larger value is suggested in Ref. \cite{Garcilazo:2007D}, namely
$a_{3/2} = -31.9$ fm, but the pertinent value for $r_{3/2}$ is not provided. 
Actually, for some potentials considered in that 
work the quartet scattering length is positive, in other words 
the $I=0$, $J^P = \frac{3}{2}^+$ state is predicted to be bound. 
Since there is no evidence for that experimently we do not consider this 
possibility here! 

\begin{figure*}[t]
\begin{center}
\includegraphics[height=72mm]{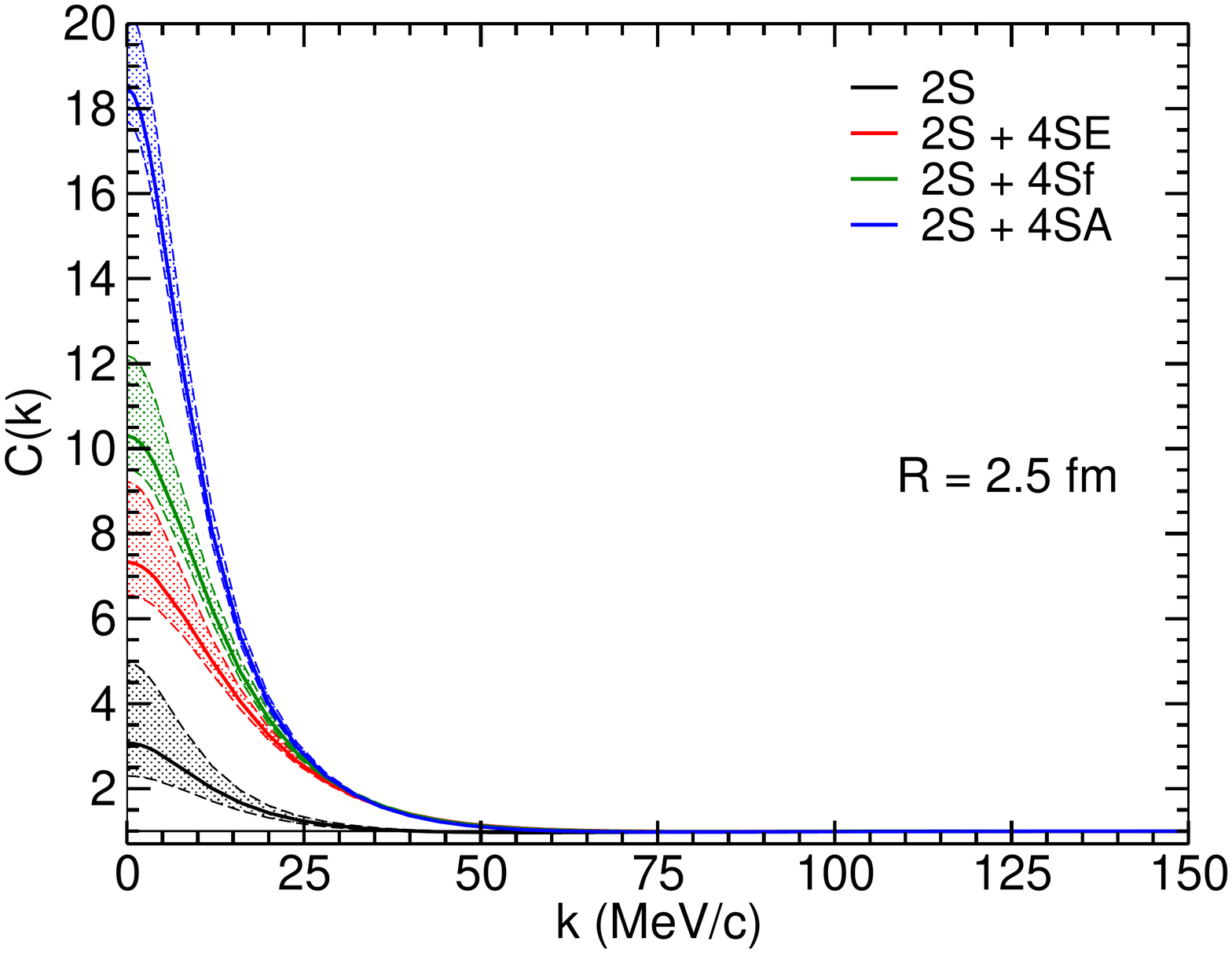}\includegraphics[height=72mm]{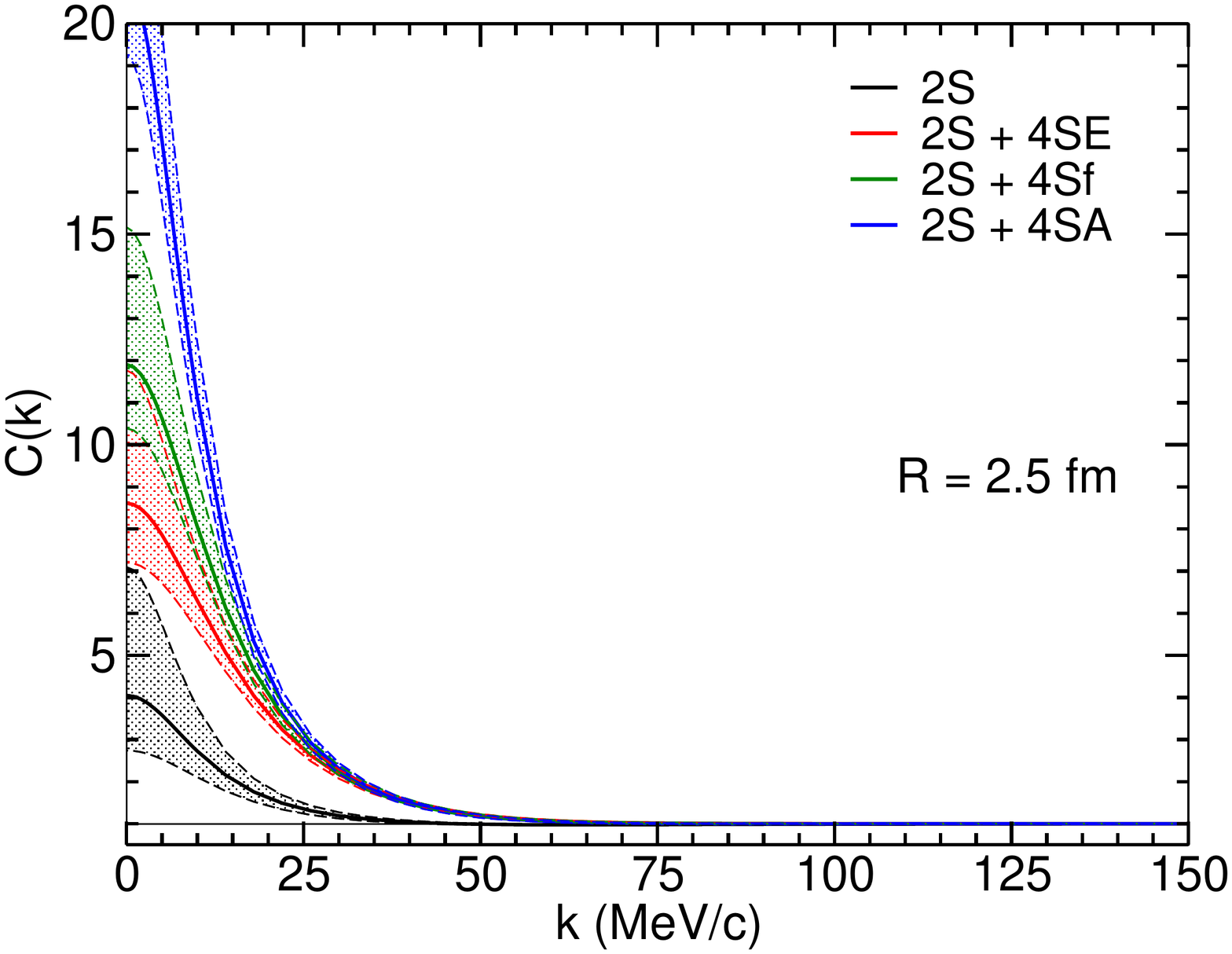}
\caption{
$\Lambda d$ correlation functions for the source size $R=2.5$ fm. Same
description of curves as in Fig. \ref{fig:Ald12}. 
}
\label{fig:Ald25}
\end{center}
\end{figure*}

\begin{figure*}[t]
\begin{center}
\includegraphics[height=72mm]{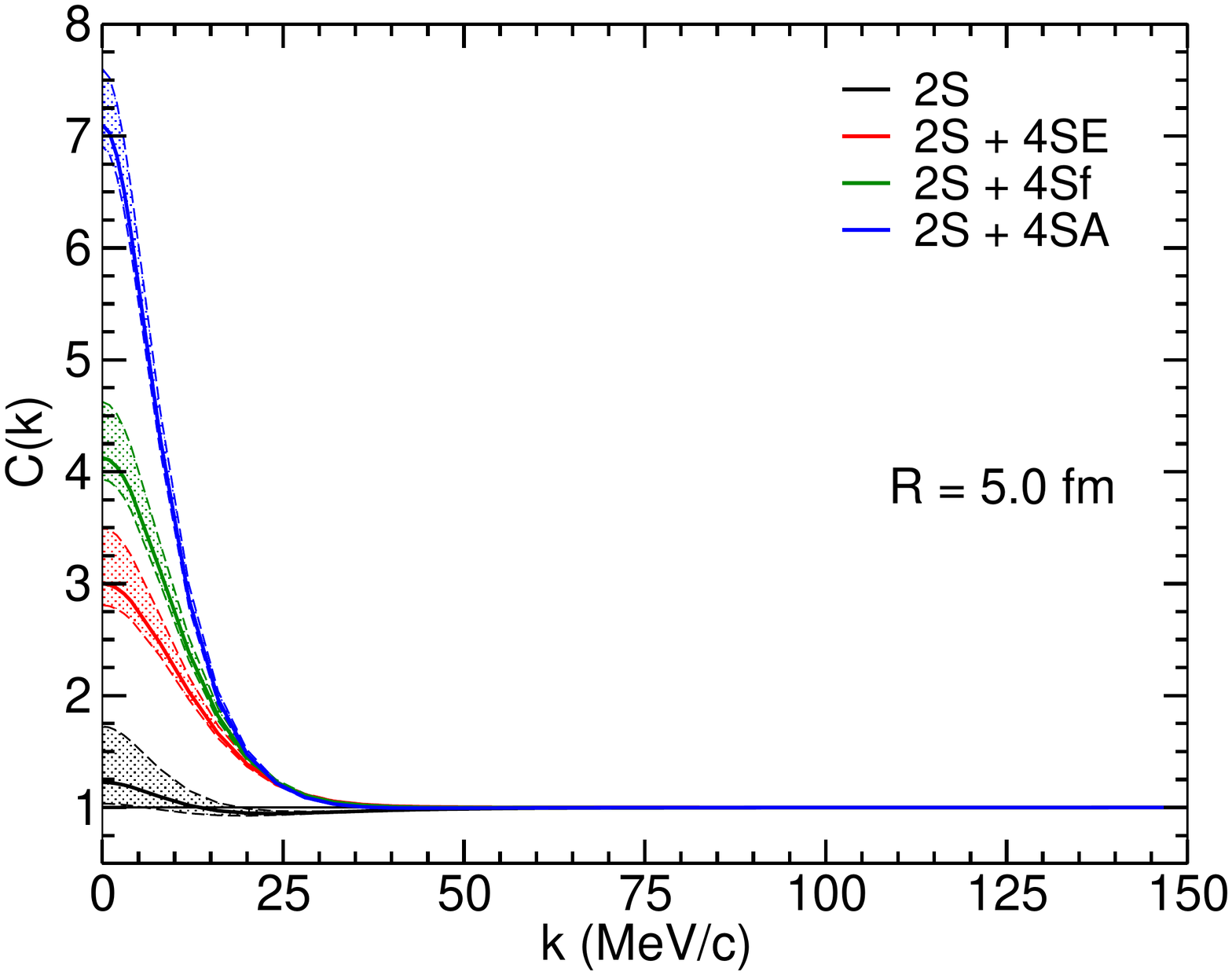}\includegraphics[height=72mm]{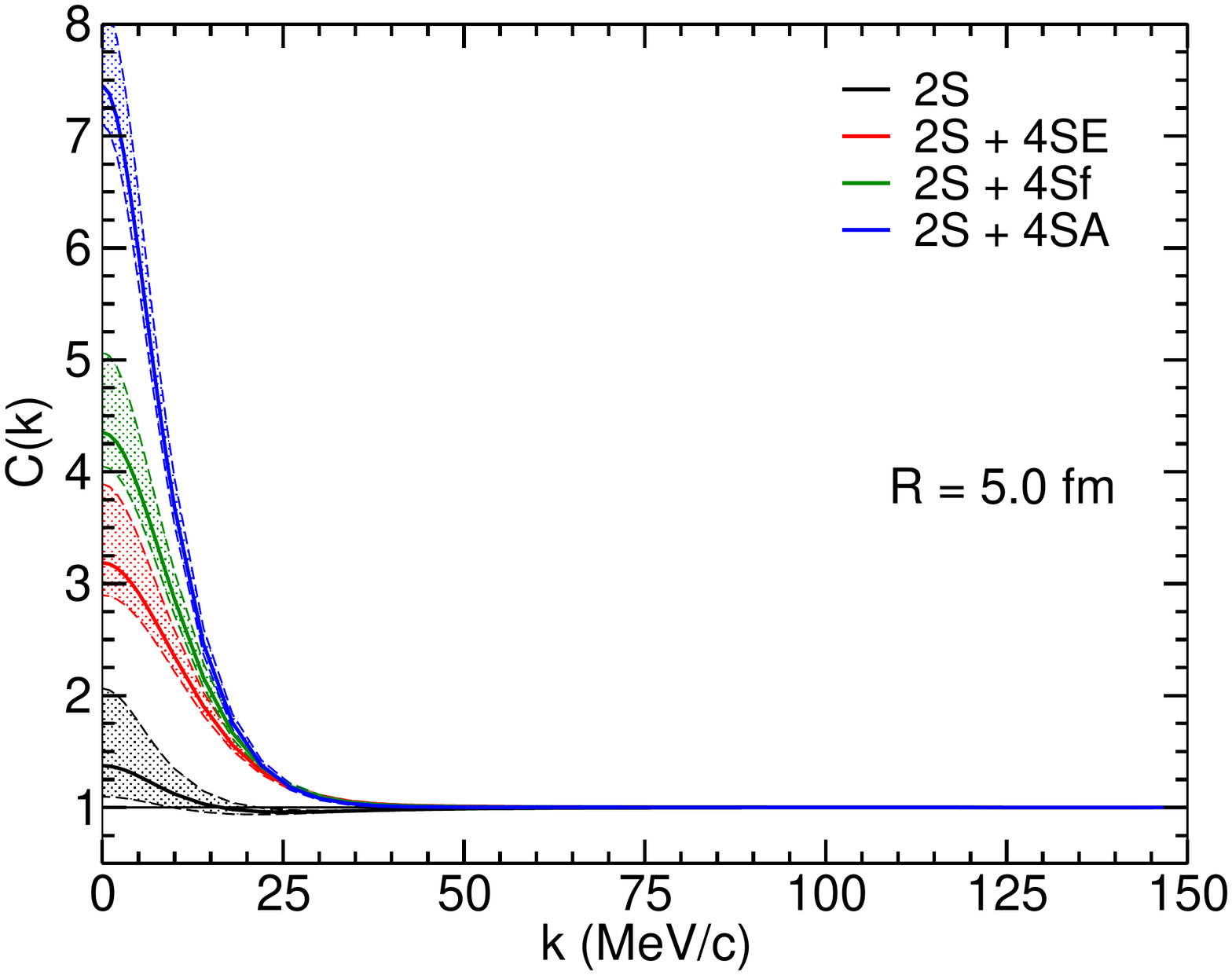}
\caption{
$\Lambda d$ correlation functions for the source size $R=5.$ fm. Same
description of curves as in Fig. \ref{fig:Ald12}. 
}
\label{fig:Ald50}
\end{center}
\end{figure*}

Our results for the $\La d$ correlation function are presented in 
Figs. \ref{fig:Ald12} (source radius $R=1.2$ fm), \ref{fig:Ald25} ($R=2.5$ fm), 
and \ref{fig:Ald50} ($R=5$ fm), respectively,
for different combinations of the doublet and quartet amplitudes. 

Evidently, the available studies suggest that the scattering lengths for the 
doublet and quartet $S$-states are both large. 
Indeed, while the former partial wave is governed by the shallow $^3_{\La}$H 
bound state, the latter is characterized by the presence of a near-threshold 
virtual state, as pointed out in Ref. \cite{Schafer:2020}.
Standard experiments allow one only to measure an average over the 
two states. As mentioned above, we make here the usual assumption 
that the weights of the spin components in the correlation function is the 
same as for free scattering ($1/3$ and $2/3$, respectively) which puts a 
somewhat larger weight on the quartet contribution.
However, equally important for the concrete results in the $\La d$ case 
is the characteristic dependence of the correlation function on the source 
radius $R$, cf. the exemplary discussion in Refs. \cite{Cho:2017,Morita:2020}. 
Specifically, for a combination of large scattering length and small $R$ the 
correlation function $C(k)$ is significantly enhanced at small values of $k$,
independently of the sign of $a$. Accordingly, one has to expect 
that in this limit the two $\La d$ partial waves yield similar effects. 
With increasing $R$ the enhancement of $C(k)$ decreases continuously in case of 
a moderately attractive interaction (negative $a$). On the other hand, if a bound state
is present 
(positive $a$), the $C(k)$ drops rather rapidly and eventually even falls below the 
nominal value of $C(k) \equiv 1$, i.e. there is a depletion as compared to the case 
without any two-particle interaction \cite{Cho:2017,Morita:2020}. 
Thus, now the two partial-wave contributions should show a rather different trend. 

After these general statements, let us discuss the results in more detail.
The predictions for $R=1.2$ fm displayed in Fig. \ref{fig:Ald12} represent
roughly the first scenario. As expected $C(k)$ is strongly enhanced at small 
momenta; 
compare this with measurements and calculations for the $\La p$ system
\cite{Acharya:2018,STAR:2006,Shapoval:2015,Acharya:2020}
where the scattering lengths are typically in the order of $2$ fm. 
Besides that, we see a sizable dependence of the correlation functions on 
the properties in the doublet wave. For the Cobis amplitude the 
contribution of the $^2S_{1/2}$ itself 
is relatively small and well separated from the results that include the 
quartet contributions based on the effective range parameters from Sch\"afer,
irrespective of the uncertainty due to the $^3_{\La}$H binding energy. 
In case of the Hammer parameters the doublet contribution is much larger 
and the uncertainties too. Indeed, now there is an overlap between the 
various results including the quartet contribution. 
A closer inspection revealed that the differences in the doublet contribution 
are primarily caused by the differences in the effective range $r$. Thus, 
a more elaborate evaluation of the doublet amplitude within EFT, beyond the 
present LO level, could presumably allow one to pin down the effective range 
more reliably. As known from studies of the $NN$ and $\La N$ systems 
within chiral EFT, LO calculations tend to underestimate the effective range. 
Of course, a reduction in the uncertainty of the $^3_{\La}$H binding 
energy would be also extremely useful. 
Anyway, under the present circumstances one has to concede that drawing reliable 
conclusions on the magnitude of the quartet amplitude from measurements 
in reactions where the source size is small is difficult. 

The second scenario discussed above is more or less realized in the results for 
$R=5$ fm presented in Fig. \ref{fig:Ald50}. Here the signal is clearly dominated 
by the quartet contribution. Those from
the doublet state are small so that the difference between the ERE
parameters from Cobis and Hammer and even the uncertainty in the $^3_{\La}$H
binding energy do not play a decisive role. Therefore, a measurement of the 
correlation function under these conditions would certainly yield valuable constraints 
on the quartet contribution and, in turn, on the corresponding scattering length. 

Note that the result for the quartet channel depends primarily on the $\La N$ 
spin-triplet interaction \cite{Miyagawa:1995,Garcilazo:2007D} and, thus, 
provides directly constraints on the latter quantity. But of course, there can 
be also contributions from 3BFs. Indeed, in the $\slashed{\pi}$EFT calculation 
of Sch\"afer et al. their influence appears to be significant \cite{Schafer:2020}. 
In that work the arising 3BF is fixed by considering the binding energy of the 
$1^+$ state of $^4_{\La}$H. 
However, one should not forget that in $\slashed{\pi}$EFT 3BFs appear 
at LO \cite{Hammer:2002,Schafer:2020,Contessi:2018}.  
We expect the situation to be different in calculations within chiral EFT where
pion exchange and, specifically, the important coupling of $\La N$ to $\Si N$
are taken into account explicitly \cite{Haidenbauer:2019}. In this scheme 3BFs 
appear first at next-to-next-to-leading order (N$^2$LO) \cite{Petschauer:2016}. 

\begin{figure*}
\begin{center}
\includegraphics[height=72mm]{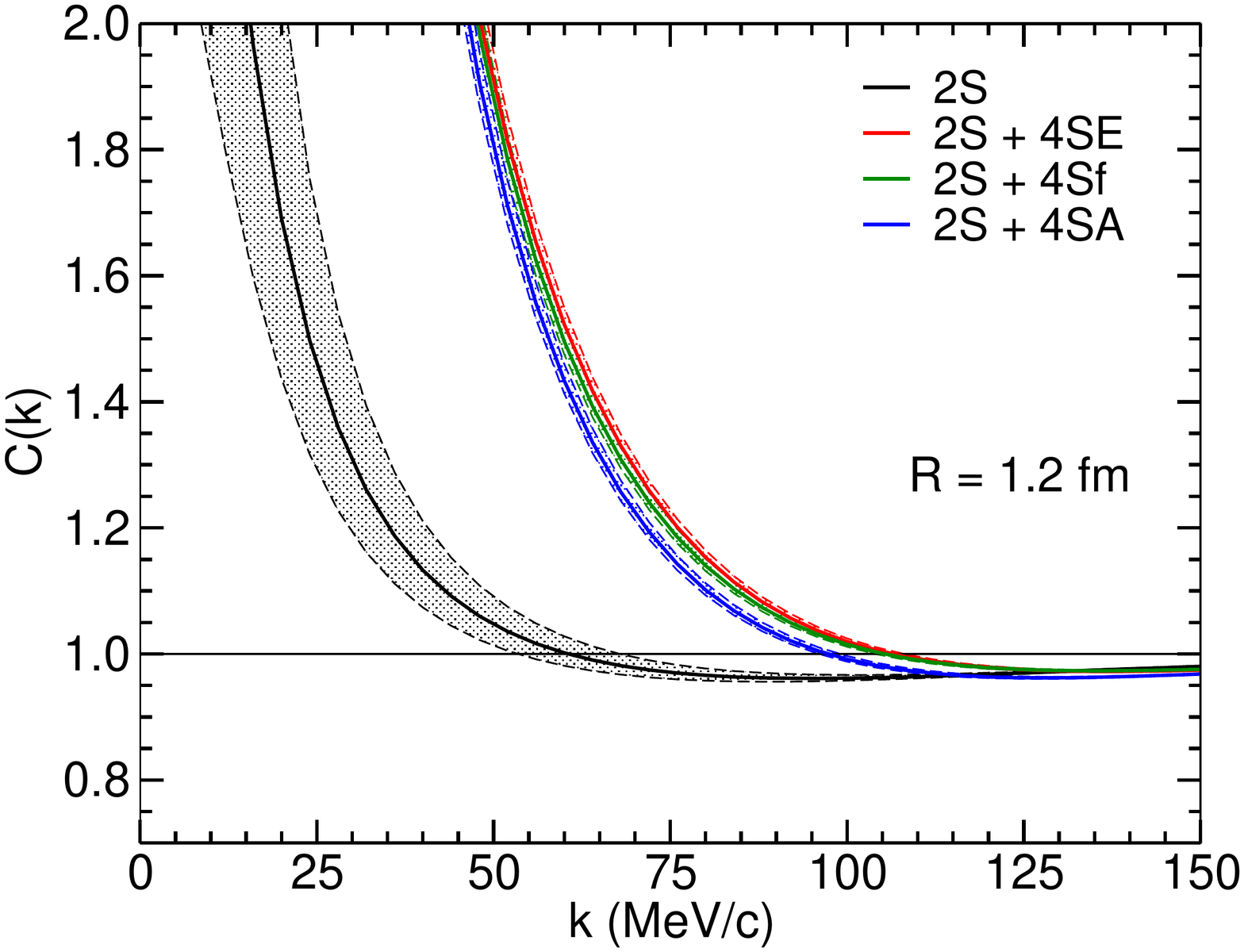}\includegraphics[height=72mm]{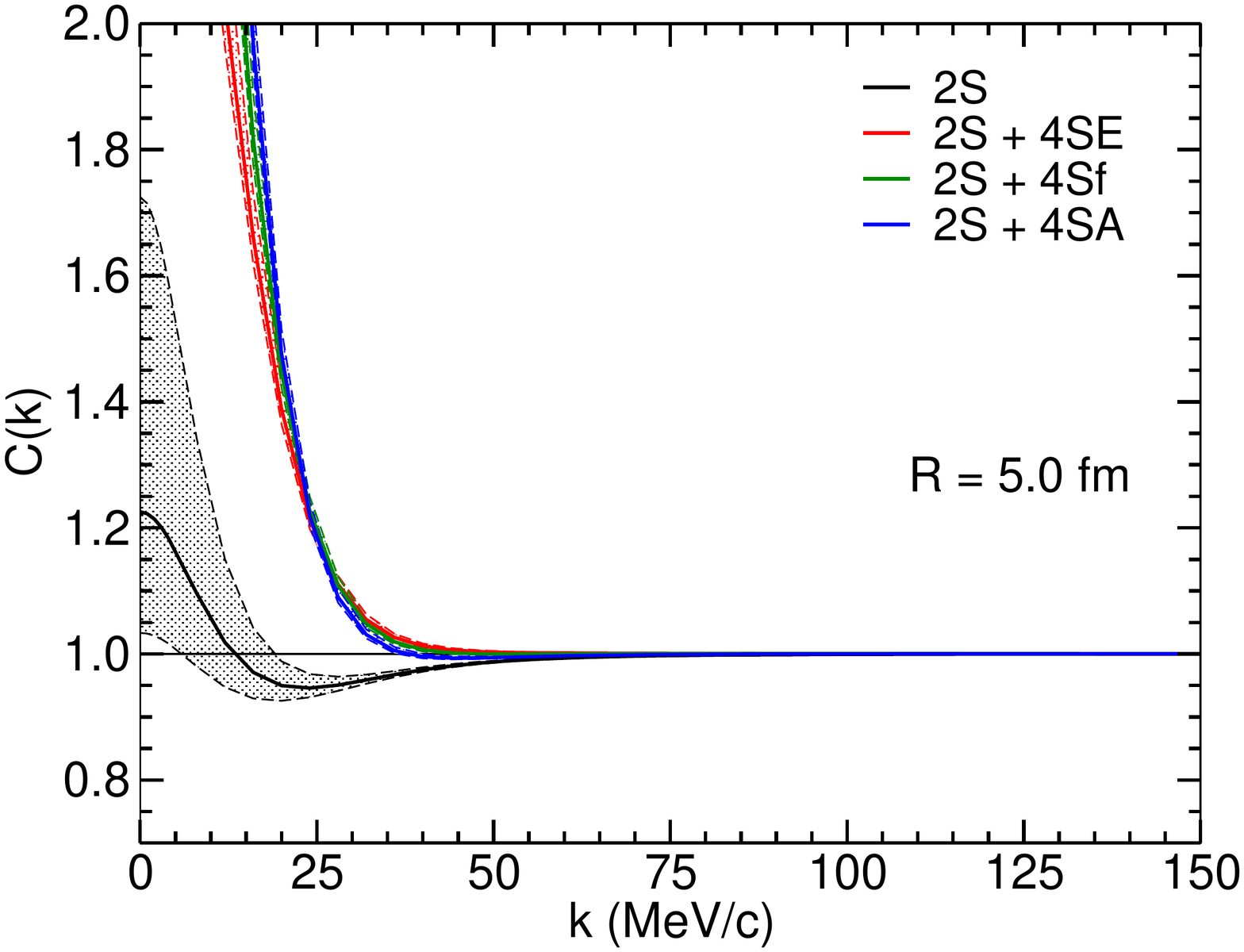}
\caption{
$\Lambda d$ correlation functions for the source sizes $R=1.2$ and $5$~fm,
based on the $^2S_{1/2}$ effective range parameters of Cobis \cite{Cobis:1997}.
Same description of curves as in Fig. \ref{fig:Ald12}. 
}
\label{fig:Ald12D}
\end{center}
\end{figure*}

Finally, we want to mention that evidence for a possibly larger hypertriton 
binding energy, $B_\La= 0.41\pm 0.12$ MeV, has been reported recently by 
the STAR Collaboration \cite{Adam:2019}. Such an energy
implies $a_{1/2}=10.2^{+1.5}_{-0.9}$~fm, asuming the Cobis value for $r_{1/2}$. 
Clearly, with that the contribution of the doublet state to $C(k)$ (and 
its uncertainty) would be drastically reduced. It is below the lower bound of the 
uncertainty shown in the figures, a situation which would be certainly beneficial 
for the determination of the quartet amplitude from a measurement of the 
$\La d$ correlation function.  
In fact, a recent calculation based on $\slashed{\pi}$EFT suggests also a distinctly 
smaller value, namely $a_{1/2}=13.8^{+3.75}_{-2.03}$~fm \cite{Hildenbrand:2019}.
However, the central value here leads to a negative result for $r_{1/2}$,
when inserted in the Bethe formula together with the $^{\La}_3$H binding 
energy of $0.13$ MeV. There is a well-known anomaly in the corresponding 
doublet state of $n\, d$ scattering which requires a modification
of the effective range function \cite{vanOers:1967}. But in case of $\La d$ 
there is no indication for an unusual behavior, judging from the plot of 
$k\cot \delta$ in Ref. \cite{Hildenbrand:2019}.
Incidentally, assuming that $r_{1/2}\equiv 0$ leads to 
$a_{1/2}=14.6^{+4.1}_{-2.2}$~fm based on Eq. (\ref{Bethe}). 

Since the correlation functions for small momenta are rather large we show selected
results on a different scale in Fig. \ref{fig:Ald12D} so that one can see the 
behavior in the region of $k=25-100$ MeV/c in detail. There is a sizable effect from
the quartet state in the region of $k = 25-50$ MeV/c. On the other hand, differences
between the different strength of the quartet amplitudes are rather difficult to 
resolve in this momentum region. 
One should be aware that in the calculation of Sch\"afer et al. all the
effective ranges are practically the same and about $3.6-3.8$ fm. It remains 
unclear whether that is a realistic range or rather a consequence of the LO treatment. 
Noticeably different values of $r_{3/2}$ could lead to stronger variations in the
momentum region of $k=25-100$ MeV/c.
In this context let us mention that the momentum corresponding to the $\La\,d$ 
breaking threshold is $k\approx 55$ MeV/c. However, judging from results for 
the elastic and total $\La\,d$ cross sections shown in the works of
Schick and collaborators \cite{Schick:1965,Schick:1971}, 
and by results for the $^4S_{3/2}$ $n\,d$ phase shift, see, e.g., Ref. \cite{Bedaque:2000},
drastic effects from the break-up are rather unlikely -- at least in the 
momentum region up to $100$ MeV/c, where $C(k)$ is noticeably different from $1$.

\section{Conclusions}

In the present paper we have investigated the potential of $\La d$ 
two-particle momentum correlation functions as an additional and 
alternative source of information on the elementary $\La N$ interaction. 
Since the present work is primarily an exploratory study, intended as an 
illustration for what can be expected from measuring $\La d$ correlations, 
it has been done on a simple technical level. Specifically, it has been performed 
within the Lednicky-Lyuboshits approach \cite{LL1} 
and by utilizing an effective range expansion for the relevant $\La d$ amplitudes.  
The effective range parameters for the two $S$-wave states that can contribute, 
the $^2S_{1/2}$ and $^4S_{3/2}$ partial waves, have been taken from results
available in the literature \cite{Cobis:1997,Hammer:2002,Schafer:2020}. 

Of specific interest is the situation in the $^4S_{3/2}$ partial wave, since presently 
there is no constraint on its properties. Thus, the main question is whether 
measurements of the correlation function could allow one to pin it down.
The $^4S_{3/2}$ state is strongly linked to the properties of the $\La N$ 
interaction in the spin triplet ($^3S_1$) state and could allow conclusions 
on the pertinent interaction strength. 
Note that the $^2S_{1/2}$ partial wave is fixed to a large extent by the 
presence of a weakly bound state, the $^\La_3$H. 

Our results suggest that measurements of the $\La d$ correlation function 
in configurations characterized by a large emitting source such as
in central heavy-ion collisisions \cite{Morita:2020} look indeed very promising. 
For large source sizes the contribution from the $^2S_{1/2}$ partial wave
is significantly suppressed, because of the presence of the hypertriton, 
and one is predominatly sensitive to the quartet state. 
In contrast, for small sizes both states contribute in a similar way and with comparable
magnitude. Given the present uncertainty in the doublet effective range parameters 
and the $^\La_3$H binding energy, respectively, conclusions are much more difficult 
to draw based on the experiment alone. 
Nonetheless, also in this case a more refined evaluation of the effective range 
parameters in the $^2S_{1/2}$ state, say based on a $\La d$ calculation utilizing 
NLO $\La N$ forces \cite{Hai:2013,Haidenbauer:2019}, and/or more accurate 
data on the $^\La_3$H binding energy could still allow one to gain further insight. 

\acknowledgments{
Work partially supported by the DFG and the NSFC through funds
provided to the Sino-German CRC 110 `Symmetries and
the  Emergence of Structure in QCD''  (DFG  grant.  no.
TRR 110).
}


\end{document}